# Electronic and plasmonic phenomena at graphene grain boundaries

Z. Fei[1], A. S. Rodin[2], W. Gannett[3,4], S. Dai[1], W. Regan[3,4], M. Wagner[1], M. K. Liu[1], A. S. McLeod[1], G. Dominguez[5,6], M. Thiemens[5], A. H. Castro Neto[7], F. Keilmann[8], A. Zettl[3,4], R. Hillenbrand[9,10], M. M. Fogler[1], D. N. Basov[1]*.

[1]Department of Physics, University of California, San Diego, La Jolla, California 92093, USA

[2]Department of Physics, Boston University, Boston, Massachusetts 02215, USA

[3]Department of Physics and Astronomy, University of California, Berkeley, California 94720, USA

[4]Materials Sciences Division, Lawrence Berkeley National Lab, Berkeley, California 94720, USA

[5]Department of Chemistry and Biochemistry, University of California, San Diego, La Jolla, California 92093, USA

[6]Department of Physics, California State University, San Marcos, California 92096, USA

[7]Graphene Research Centre and Department of Physics, National University of Singapore, 117542, Singapore

[8]Ludwig-Maximilians-Universität and Center for Nanoscience, 80539 München, Germany

[9]CiC nanoGUNE Consolider, 20018 Donostia-San Sebastián, Spain

[10]IKERBASQUE, Basque Foundation for Science, 48011 Bilbao, Spain

*Correspondence to: dbasov@physics.ucsd.edu

**Graphene[1], a two-dimensional honeycomb lattice of carbon atoms, is of great interest in (opto)electronics[2,3] and plasmonics[4-11] and can be obtained by means of diverse fabrication techniques, among which chemical vapor deposition (CVD) is one of the most promising for technological applications[12]. The electronic and mechanical properties of CVD-grown graphene depend in large part on the characteristics of the grain boundaries[13-19]. However, the physical properties of these grain boundaries remain challenging to characterize directly and conveniently[15-23]. Here, we show that it is possible to visualize and investigate the grain boundaries in CVD-grown graphene using an infrared nano-imaging technique. We harness surface plasmons that are reflected and scattered by the graphene grain boundaries, thus causing plasmon interference. By recording and analyzing the interference patterns, we can map grain boundaries for a large area CVD-grown graphene film and probe the electronic properties of individual grain boundaries. Quantitative analysis reveals that grain boundaries form electronic barriers that obstruct both electrical transport and plasmon propagation. The effective width of these barriers (~10-20 nm) depends on the electronic screening**

and it is on the order of the Fermi wavelength of graphene. These results uncover a microscopic mechanism that is responsible for the low electron mobility observed in CVD-grown graphene, and suggest the possibility of using electronic barriers to realize tunable plasmon reflectors and phase retarders in future graphene-based plasmonic circuits.

Our imaging technique, which we refer to as 'scanning plasmon interferometery', is implemented in a setting of an antenna-based infrared (IR) nanoscope[6-8]. A schematic diagram of the scanning plasmon interferometry technique is shown in Fig. 1a. Infrared light focused on a metalized tip of an atomic force microscope (AFM) generates a strong localized field around the sharp tip apex, analogous to a "lightning-rod" effect[24]. This concentrated electric field launches circular SPs around the tip (pink circles in Fig. 1a). The process is controlled by two experimental parameters: the wavelength of light $\lambda_{IR}$ and the curvature radius of the tip $R$. In order to efficiently launch SPs on our highly doped graphene films, we chose IR light with $\lambda_{IR}$ close to 10 μm and AFM tips with $R \approx$ 25 nm (Methods). The experimental observable of the scanning plasmon interferometry is the scattering amplitude $s$ that is collected simultaneously with AFM topography.

Before analyzing the GBs, we first discuss a crack-type line defect with a geometric width of ~10 nm, thus visible in the AFM topography (blue arrows in Fig. 1b). The corresponding scanning plasmon interferometry image is displayed in Fig. 1c, where we plot the scattering amplitude $s$ at $\lambda_{IR}$ = 11.3 μm. The scattering signal shows bright twin fringes running along this line defect. In the same field of view, we also observed a region of double-layer graphene (blue dashed loop) and a microscopic line structure (green shaded region) in Fig. 1b. All these features are commonly found in CVD graphene[12] (Fig. S1a). The bright circular fringes are observed near the edge of the double-layer region (Fig. 1c). By tuning $\lambda_{IR}$ from 11.3 μm (Fig. 1c) to 10.5 μm (Fig. 1d), the fringe widths of both types of fringes show evident $\lambda_{IR}$-dependence, which is consistent with the plasmonic origin of these patterns[7,8]. Note that the scattering amplitude in all our scanning plasmon interferometry images is normalized to that of a sample region where no fringes exist (e.g. the green square in Fig. 1c).

In previous studies[7,8], plasmon fringes with a width of half the plasmon wavelength $\lambda_p/2$ were observed close to the edge of graphene microcrystals. In order to validate the plasmonic origin of the fringes found here, we plot in Fig. 1f the width of the twin fringes (circles) as a function of $\lambda_{IR}$. In the same diagram we also show a theoretical cacluation (see Methods for details). The agreement between the experimental data and the calculated curve confirms that the bright fringes at the line defects are of the plasmonic origin in close analogy with the oscillations of the scattering amplitude at the edges of graphene. In either case, the near-field signal is formed by a standing wave with the periodicity $\lambda_p/2$ produced by the interference between the tip-launched and reflected plasmons[7,8].

We observed twin fringes not only close to the cracks but also near other types of line defects that we identified as wrinkles and grain-overlaps based on the AFM topography (Fig. S2). But the most prevailing line defects are grain boundaries (schematically illustrated in Fig. 1a with a red line). As a rule, GBs are of the atomic length scale thus are invisible in typical AFM scans (Fig. 2a). Yet GBs were vividly visualized by scanning plasmon interferometry producing characteristic twin fringes (Fig. 2b,d). We examined the $\lambda_{IR}$-dependence of the fringe width and found that it is in agreement with the theoretical calculation (red circles in Fig. 1f). This latter finding attests to the plasmonic origin of the scanning plasmon interferometry signal at GBs.

So far we discuss mainly the fringe width that is a direct measure of $\lambda_p$. Yet another important parameter is the separation between twin fringes $D_{TF}$ (Fig. 1e). For GBs, $D_{TF}$ can be written as $D_{TF} \approx (-\delta/2\pi)\lambda_p$, where $\delta$ is the plasmon phase shift upon reflection off a grain boundary set to vary within [-2$\pi$, 0] (Supplementary equation (S19)). Therefore, for a non-zero constant $\delta$, the magnitude of $D_{TF}$ is proportional to $\lambda_p$, which was indeed confirmed by our experiment (Fig. 1f). Our data indicated that $D_{TF}$ roughly equals to $1/2\lambda_p$ for all GBs, and therefore $\delta$ is close to $-\pi$. Note that the parameter $\delta$ is not solely determined by the response of our graphene samples. The AFM tip also plays an important role here. As detailed in Section 4 of the Supplementary Information, it is convenient to write $\delta$ as $\delta = \delta_{sp} + \delta_t$, where $\delta_{sp}$ is the plasmon phase shift without tip coupling to graphene, and $\delta_t$ is a tip-dependent parameter that is around $-(0.5\pm0.1)\pi$ based on our numerical modeling (Supplementary equation (S19)).

The above analysis for $D_{TF}$ holds true also for other types of line defects with geometric features much smaller than $\lambda_p$, such as the crack shown in Fig. 1b. Nevertheless, for line defects such as wrinkles and grain-overlaps (Fig. S2), the twin fringes are strongly affected by their geometric form. As detailed in Section 2 of the Supplementary Information, these two types of line defects generate twin fringes with considerable variations of $D_{TF}$ governed by the details of a particular defect. A unique feature of GBs and grain-overlaps is that they together form a network of closed regions (grains) spanning over the entire graphene film (Figs. 2e & S3). In contrast, cracks and wrinkles are sporadic and discontinuous. From Fig. 2e, we were able to measure the average grain size (3-5μm) of our film, in agreement with reports for graphene prepared under identical conditions[21].

In order to gain quantitative understanding of the twin fringes in our scanning plasmon interferometry images, we performed numerical modeling that takes into account all the experimental details. In our modeling, we assumed that GBs locally modify the plasmon wavelength $\lambda_p$ and damping rate $\gamma_p$. Here, $\gamma_p$ is defined by the ratio between the imaginary and real parts of the plasmon wavevector $q_p \equiv \frac{2\pi}{\lambda_p}(1+i\gamma_p)$. We found that the profiles of $\lambda_p(x)$ and $\gamma_p(x)$ displayed in Fig. 2f produce an accurate fit of the experimental data taken at multiple $\lambda_{IR}$ in the range of 10.7–11.3 μm (Figs. 2c, Supplementary Fig. S7). Details of the modeling are given in Section 5 of the Supplementary Information. The fact that the single set of parameters fits the totality of fringe profiles indicates that the choice of these parameters is quite robust. For example, an assumption of a dip in $\lambda_p(x)$ as opposed to a peak at the GB would almost double $D_{TF}$ (see Fig. S5a and following paragraphs). We remark that strong scattering quantified with $\gamma_p$ in concert with the enhancement of $\lambda_p$ at the GB is needed to reproduce the line shape of the twin fringes.

We now discuss some of the implications of our modeling. According to the plasmon dispersion equation (Methods), $\lambda_p$ is roughly proportional to $E_F$. In turn, $E_F$ scales as a square root of the carrier density $n$. Thus our results imply that our graphene film tends to be heavily doped with $n \approx 4 \times 10^{13}$ cm$^{-2}$ at the GBs, corresponding to 0.021 holes per unit cell. This is expected since GBs are lattice defects that favor molecule adsorptions at ambient conditions[25,26]. The role of defects in enhancing doping due to molecule adsorption has been extensively studied before[27,28]. In contrast, under ultra-high-vacuum conditions, where molecule adsorption is significantly reduced, graphene films are close to the charge neutrality point and GBs perturbed the electronic properties of graphene in a totally different way as confirmed by scanning tunnel microscopy studies[18]. The plasmon damping rate depends on the carrier scattering rate of graphene $\tau^{-1}$: $\gamma_p \approx 0.05 + (\omega\tau)^{-1}$ (Eq. S21). Therefore, the lineform of $\gamma_p(x)$ inferred from modeling implies that charge carriers experience enhanced scattering close to the GBs. We speculate that this effect may be caused by the coulomb scattering due to the charges at the GBs. Furthermore, modeling indicates that GBs perturb electronic properties over a length scale of the order of 20 nm. A wider effective width compared to the geometric width is in fact an outcome of electron screening of the charged GBs[29]. Indeed, the charge screening length is estimated to be in the order of Fermi wavelength, roughly 10 nm in our doping range, consistent with our experimental findings.

Based on the $\lambda_p(x)$ and $\gamma_p(x)$ profiles in Fig. 2f, we can calculate the $E_F(x)$ and $\tau^{-1}(x)$ profiles across GBs. These latter parameters allow us to infer the DC conductivity $\sigma_{DC}$ of graphene (inset of Fig. 2c) with a standard formula[11]: $\sigma_{DC} \approx \frac{2e^2}{h} \frac{E_F}{\hbar\tau^{-1}}$. This equation is obtained by assuming weak frequency dependence of $\tau^{-1}$ that is valid when coulomb scattering dominates[11]. Although the increase of the $E_F$ near the GBs would normally boost $\sigma_{DC}$, this expected trend is overwhelmed by the increase in $\tau^{-1}$. The net effect for GBs is to significantly reduce the local $\sigma_{DC}$ of graphene.

Finally, we wish to point out that the plasmon reflection off GBs can be described by a reflection coefficient $r_{sp}$. By solving analytically the problem of SPs scattering by GBs, we were able to obtain a formula: $r_{sp} \approx iW_{eff}\Delta q_p$ (Eq. S14), where $\Delta q_p$ is the relative change of plasmon wavevector due to GBs and $W_{eff}$ is the effective width. Based on $r_{sp}$, we were able to estimate both the plasmon reflection probability $|r_{sp}|^2$ and the phase shift $\delta_{sp}$=arg($r_{sp}$). The former is closely related to the fringe intensity, and the latter determines $D_{TF}$ as discussed above. Calculations based on the $\lambda_p$ and $\gamma_p$ profiles of the GB (Fig. 2f) yield $|r_{sp}|^2 \approx 8\%$ and $\delta_{sp} \approx -0.6\pi$. The 8% reflectivity is remarkably high. Such a strong reflection is due to the extended effective width of the electronic perturbation induced by the GBs (Fig. 2f). A phase shift of $\delta_{sp} \approx -0.6\pi$ is an outcome of higher doping at the GB. If one switches the GB to a lower doping, $\delta_{sp}$ will undergo a "$\pi$" phase shift and become -1.6$\pi$, resulting in a dramatic increase of $D_{TF}$ away from the experimental value (Fig. S5a). The above analysis indicates that $|r_{sp}|^2$ and $\delta_{sp}$ are sensitive to the doping of the plasmon reflector. Therefore both of these parameters governing plasmon propagation can be conveniently tunable by common electronic means, e.g. electrostatic gating.

Our work provided for the first time unambiguous experimental evidence of novel plasmonic effects originating from plasmon reflection at GBs in CVD graphene. The scanning plasmon interferometry technique, aided with modeling, is a comprehensive method capable of mapping and probing the electronic properties of GBs. This method can be applied to nano-characterization of plasmonic materials beyond graphene, where GBs also play important roles in the plasmonic effects[30]. Moreover, our work provides guidelines to designing tunable electronic barriers that would realize reconfigurable plasmon reflectors[4] and phase retarders: a milestone towards graphene-based plasmonic circuits.

## Methods
### Samples
Our graphene films were grown on copper foils using a two-step low pressure CVD method[12], and then transferred to silicon wafers with 300 nm $SiO_2$ layer on top. All experiments were performed under ambient conditions and in an atmospheric environment. The graphene films were unintentionally hole-doped with a carrier density of about $1.0 \times 10^{13}$ cm$^{-2}$ corresponding to a Fermi energy $E_F$ of 0.37 eV. Such high doping is due to the $SiO_2$ substrate, as well as molecule adsorption in the air atmosphere[26,27]. The doping level was inferred from our Raman and near-field gating experiments (Supplementary Section S1).

### Experimental apparatus
The scanning plasmon interferometry experiments introduced in the main text were performed at UCSD using a scattering-type scanning near-field optical microscope (s-SNOM)[24]. Our s-SNOM is a commercial system (neaspec.com) equipped with mid-IR quantum cascade lasers (daylightsolutions.com) and $CO_2$ lasers (accesslaser.com) covering a wavelength range of 9.5–11.3 μm. The s-SNOM equipped with a pseudo-heterodyne interferometric detection module is based on an atomic force microscope (AFM) operating in the tapping mode with a tapping frequency around 270 kHz. The output signal of s-SNOM utilized in this work is the scattering amplitude $s$ demodulated at $n^{th}$ harmonics of the tapping frequency ($n = 2$ in the current work).

In order to efficiently couple IR light to the graphene plasmons, an AFM tip with a radius $R \approx 25$ nm was chosen as our near-field probe. This scheme allowed us to overcome the notorious "momentum mismatch" between plasmons and photons. As detailed in ref. 13, the momenta-coupling function has a bell-shaped momenta distribution that peaks at $q \sim 1/R$. For a typical CVD graphene film on the $SiO_2$ substrate, the momentum of IR plasmons at ambient conditions is between $3 - 6 \times 10^5$ cm$^{-1}$. Therefore the optimum tip radius for exciting SPs of graphene in our frequency range is about 20-30 nm.

### Evaluating the plasmon dispersion in graphene
The plasmon dispersion equation of graphene[7,11] at the interface between air and $SiO_2$ substrate with dielectric function $\varepsilon_{sub}(\omega)$ is given as $q_p = \dfrac{i2\omega\varepsilon_0\kappa(\omega)}{\sigma(\omega)}$, where $\omega = 2\pi c/\lambda_{IR}$ is the IR excitation frequency, $\kappa(\omega) = [1+\varepsilon_{sub}(\omega)]/2$ is the effective dielectric function of the environment for graphene, $\sigma(\omega)$ is the optical conductivity of graphene. The plasmon wavelength $\lambda_p$ of graphene can be obtained with $\lambda_p=2\pi/\text{Re}(q_p)$. The optical conductivity we used to calculate the plasmon wavelength ($\times 1/2$) in Fig. 1f was obtained from the random phase approximation method[6,7]. We find an excellent agreement between the experimental data and calculations of $1/2\lambda_p$ assuming a Fermi energy $E_F \approx$ 0.37 eV that is in accord with our Raman measurements.

Alternatively, one can use a Drude formula that is valid at a limit of long wavelength and low frequency: $\sigma(\omega) = i\frac{e^2}{\pi\hbar^2}\frac{E_F}{\omega + i\tau^{-1}}$, where $e$ is the elementary charge, $\hbar$ is the reduced Plank constant, and $\tau^{-1}$ is the charge scattering rate in graphene. In this case, plasmon wavelength $\lambda_p$ adopts an analytic form: $\lambda_p \approx \frac{e^2 E_F \lambda_{IR}^2}{h^2 c^2 \varepsilon_0 \text{Re}\kappa}$.

**Acknowledgments:**
Authors acknowledge support from ONR. The development of scanning plasmon interferometry is supported by DOE-BES. G.D. and M.T. were supported by NASA grant NNX11AF24G. M.F. is supported by UCOP and NSF PHY11-25915. A.H.C.N. acknowledges NRF-CRP grant R-144-000-295-281. M.W. thanks the Alexander von Humboldt Foundation for financial support. R.H. acknowledges the ERC Starting Grant No. 258461. A.S.M is supported by a U.S. Dept. of Energy Office of Science Graduate Fellowship.

**Author contributions**
All authors were involved in designing the research, performing the research, and writing the paper.

**Additional information**
Supplementary information is available in the online version of the paper. Reprints and permission information is available online at www.nature.com/reprints. Correspondence and requests for materials should be addressed to D.N.B.

**Competing financial interests**
F.K. and R.H. are cofounders of Neaspec, producer of the s-SNOM apparatus used in this study.

**References**

1. Geim, A. K. & Novoselov, K. S. The rise of graphene. *Nature Mater.* **6**, 183-191 (2007).
2. Castro Neto, A. H., Guinea, F., Peres, N. M. R., Novoselov, K. S. & Geim, A. K. The electronic properties of graphene. *Rev. Mod. Phys.* **81**, 109-162 (2009).

3. Bonaccorso, F., Sun, Z., Hasan, T. & Ferrari, A. C. Graphene photonics and optoelectronics. *Nature Photon.* **4**, 611-622 (2010).

4. Vakil, A. & Engheta, N. Transformation optics using graphene. *Science* **332**, 1291-1294 (2011).

5. Ju, L. et al. Graphene plasmonics for tunable terahertz metamaterials. *Nature Nanotech.* **6**, 630-634 (2011).

6. Fei, Z. et al. Infrared nanoscopy of Dirac plasmons at the graphene-$SiO_2$ interface. *Nano Lett.* **11**, 4701-4705 (2011).

7. Fei, Z. et al. Gate-tuning of graphene plasmons revealed by infrared nano-imaging. *Nature* **487**, 82-85 (2012).

8. Chen, J. et al. Optical nano-imaging of gate-tunable graphene plasmons. *Nature* **487**, 77-81 (2012).

9. Yan, H. et al. Tunable infrared plasmonic devices using graphene/insulator stacks. *Nature Nanotech.* **7**, 330-334 (2012).

10. Grigorenko, A. N., Polini, M. & Novoselov, K. S. Graphene plasmonics. Nature Photon. **6**, 749-758 (2012).

11. Jablan, M., Buljan, H. & Slojačić, M. Plasmonics in graphene at infrared frequencies. *Phys. Rev. B* **80**, 245435 (2009).

12. Li, X. et al. Large-area synthesis of high-quality and uniform graphene films on copper foils. *Science* **324**, 1312-1314 (2009).

13. Grantab, R., Shenoy, V. B. & Ruoff, R. S. Anomalous strength characteristics of tilt grain boundaries in graphene. *Science* **330**, 946-948 (2010).

14. Wei, Y. et al. The nature of strength enhancement and weakening by pentagon-heptagon defects in graphene. *Nature Mater.* **11**, 759-763 (2012).

15. Yu, Q. et al. Control and characterization of individual grains and grain boundaries in graphene grown by chemical vapour deposition. *Nature Mater.* **10**, 443-449 (2011).

16. Song, H. S. et al. Origin of the relatively low transport mobility of graphene grown through chemical vapor deposition. *Sci. Rep.* **2**, 337 (2012).

17. Tsen, A. W. et al. Tailoring electrical transport across grain boundaries in polycrystalline graphene. *Science* **336**, 1143-1146 (2012).

18. Koepke, J. C. et al. Atomic-scale evidence for potential barriers and strong carrier scattering at graphene grain boundaries: a scanning tunneling microscopy study. *ACS Nano.* **7**, 75-86 (2013).

19. Tapasztó, L. et al. Mapping the electronic properties of individual graphene grain boundaries. *Appl. Phys. Lett.* **100**, 053114 (2012).

20. Huang, P. Y. et al. Grains and grain boundaries in single-layer graphene atomic patchwork quilts. *Nature* **469**, 389-392 (2011).

21. Kim, K. et al. Grain boundary mapping in polycrystalline graphene. *ACS Nano* **5**, 2142-2146 (2011).

22. Duong, D. L. et al. Probing graphene grain boundaries with optical microscopy. *Nature* **490**, 235-239 (2012).

23. Kim, D. W., Kim, Y. H., Jeong, H. S. & Jung, H. T. Direct visualization of large-area graphene domains and boundaries by optical birefringency. *Nature Nanotech.* **7**, 29-34 (2011).

24. Atkin, J. M., Berweger, S., Jones, A. C. & Raschke, M. B. Nano-optical imaging and spectroscopy of order, phases, and domains in complex solids. *Adv. Phys.* **61**, 745-842 (2012).

25. An, J. et al. Domain (grain) boundaries and evidence of "twinlike" structures in chemically vapor deposited grown graphene. *ACS Nano* **5**, 2433-2439 (2011).

26. Ryu, S. et al. Atmospheric oxygen binding and hole doping in deformed graphene on a $SiO_2$ substrate. *Nano Lett.* **10**, 4944-4951 (2010).

27. Das, A., Chakraborty, B. & Sood, A. K. Raman spectroscopy of graphene on different substrates and influence of defects. *Bull. Mater. Sci.* **31**, 579-584 (2008).

28. Kim, D. C. et al. The structural and electrical evolution of graphene by oxygen plasma-induced disorder. *Nanotechnology* **20**, 375703 (2009).

29. Radchenko, T. M., Shylau, A. A. & Zozoulenko, I. V. Effect of charged line defects on conductivity in graphene: numerical Kubo and analytical Boltzmann approaches. *Phys. Rev. B* **87**, 195448 (2013).

30. Y-J. Lu et al. Plasmonic nanolaser using epitaxially grown silver film. *Science* **337**, 450 (2012).

**Figure Captions:**

**Figure 1 | Probing CVD graphene with scanning plasmon interferometry. a**, Illustration of the scanning plasmon interferometry principle. The AFM tip (silver cone) illuminated with infrared (IR) light (purple cone) launches surface plasmon waves (pink circles) in graphene. These waves are partially reflected by the line defect (red line) thus causing interference between the launched and back-reflected plasmonic waves. **b,** AFM topography of CVD graphene revealing a crack-type line defect (blue arrows), double-layer graphene region (blue dashed loop), and a microscopic line structure (green shaded region). **c, d** Scanning plasmon interferometry images taken simultaneously with the AFM topography in **b** at IR wavelength $\lambda_{IR}$=11.3 µm and 10.5 µm, respectively. **e**, Line profiles taken along the white dashed lines in **c** and **d**. Here we also illustrate, for the 11.3 µm case, a protocol to extract the fringe width (FW) and the separation between the twin fringes $D_{TF}$. **f**, Evolution of fringe width (circles) and $D_{TF}$ (triangles) with $\lambda_{IR}$ for the crack in Fig. 1b (blue) and the grain boundary (GB) in Fig. 2 (red). The black solid line is a theoretical result for the magnitude of $1/2\lambda_p$ assuming the Fermi energy $E_F \approx 0.37$eV (Methods). Note that $\lambda_p$ decreases rapidly for $\lambda_{IR}$ < 10 µm: a consequence of the plasmon coupling to the surface optical phonon of $SiO_2$. The data range for GBs is narrower than that that of the crack due to the fact that GB is a less efficient plasmon reflector compared to the crack. Scanning plasmon interferometry images **c** and **d** show the normalized amplitude *s* of the nano-optic signal as described in the text. Scale bars in **b-d** are all 200 nm.

**Figure 2 | Grain boundaries observed in CVD graphene films. a,** Topography image of graphene containing GBs. **b,** Scanning plasmon interferometry image simultaneously taken with **a** at $\lambda_{IR}$=11.3 µm revealing GBs. **c,** Experimental (black squares) and modeled (red curves) twin fringe profiles. The experimental profile is taken along the dashed line in **b**. The inset shows the profile of DC conductivity inferred from modeling. **d,** Scanning plasmon interferometry image of the same sample area of **b** taken at $\lambda_{IR}$=10.7 µm. **e,** A larger-area scan of a typical sample revealing multiple grains (displayed with different false colors) defined by the twin fringes due to GBs and grain-overlaps. Details of line defects arrangements in this map are given in Fig. S3. **f,** The profiles of plasmon wavelength $\lambda_p$ and damping rate $\gamma_p$ used for modeling the fringe profiles of the GB shown in **c** and Fig. S7. Scale bars in **a, b, d** are 200 nm, and the scale bar in **e** is 1 µm.

**Figure 1**

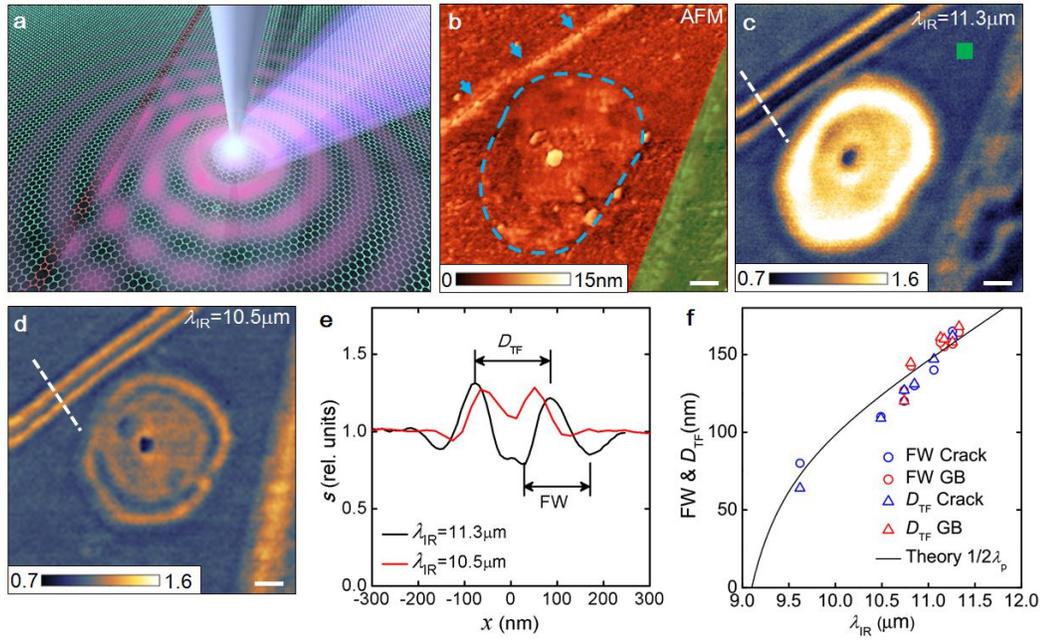

**Figure 2**

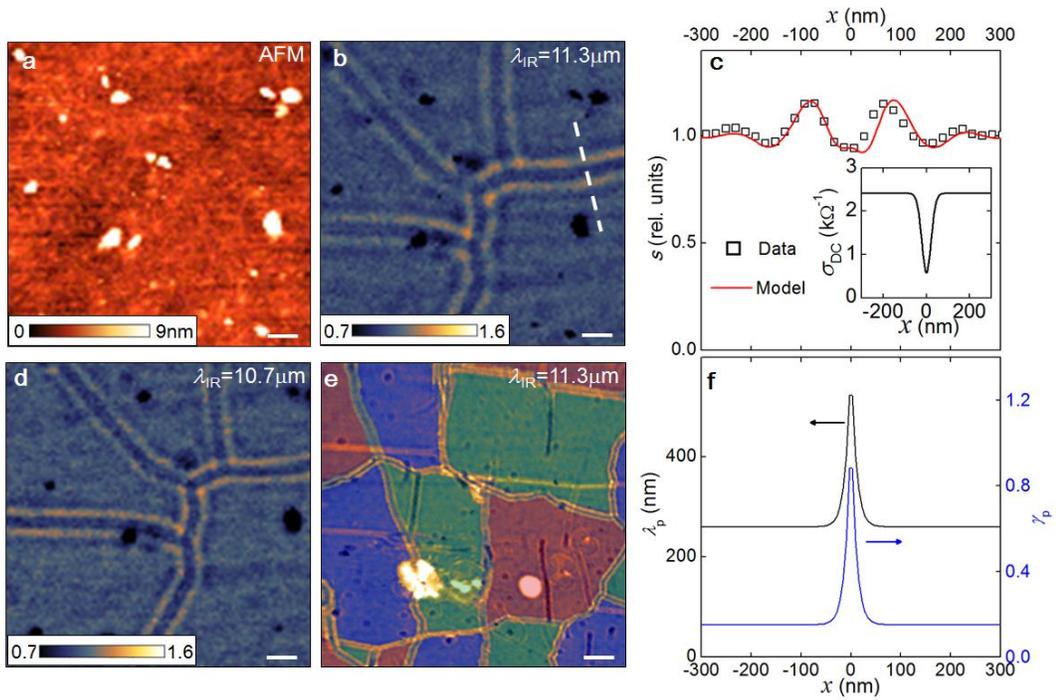

# Supplementary Information for

# "Electronic and plasmonic phenomena at grain boundaries in chemical vapor deposited graphene"

Z. Fei[1], A. S. Rodin[2], W. Gannett[3,4], S. Dai[1], W. Regan[3,4], M. Wagner[1], M. K. Liu[1], A. S. McLeod[1], G. Dominguez[6], M. Thiemens[5], A. H. Castro-Neto[7], F. Keilmann[8], A. Zettl[3,4], R. Hillenbrand[9,10], M. M. Fogler[1], D. N. Basov[1]*.

[1]Department of Physics, University of California, San Diego, La Jolla, California 92093, USA

[2]Department of Physics, Boston University, Boston, Massachusetts 02215, USA

[3]Department of Physics and Astronomy, University of California, Berkeley, California 94720, USA

[4]Materials Sciences Division, Lawrence Berkeley National Lab, Berkeley, California 94720, USA

[5]Department of Chemistry and Biochemistry, University of California, San Diego, La Jolla, California 92093, USA

[6]Department of Physics, California State University, San Marcos, California 92096, USA

[7]Graphene Research Centre and Department of Physics, National University of Singapore, 117542, Singapore

[8]Ludwig-Maximilians-Universität and Center for Nanoscience, 80539 München, Germany

[9]CiC nanoGUNE Consolider, 20018 Donostia-San Sebastián, Spain

[10]IKERBASQUE, Basque Foundation for Science, 48011 Bilbao, Spain

*Correspondence to: dbasov@physics.ucsd.edu

## 1. CVD graphene fabrication and characterization

Our graphene films were gown on copper foil using a two-step low pressure chemical vapor deposition (CVD) method as described in Ref. 1, and then transferred to $SiO_2$/Si wafers. A typical image taken with optical microscope is shown in Fig. S1a, where one can see that our CVD graphene film is predominantly single layer graphene. In addition, there are sporadic dark spots (green arrow) and lines (blue arrows) dispersed inside the film: a common occurrence in CVD graphene films[1]. These dark spots are the regions of two- or three-layer graphene whereas dark lines are microscopic line structures. We remark that these line structures are of the microscopic length scale, orders of magnitude wider than the nanoscale line defects investigated in this work. The double-layer region in Fig. 1b of the main text marked with blue dashed loop is one of these dark spots, while the green shaded region in Fig. 1b of the main text is one of these microscopic line structures.

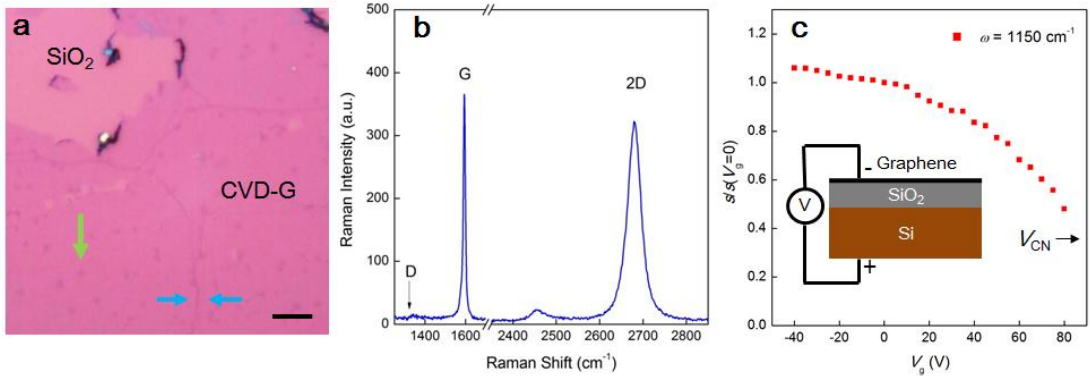

**Figure S1. | Optical and Raman characterization of CVD graphene. a**, A typical optical image of our graphene film. CVD-G represents the CVD graphene film. Green arrow marks a dark spot and blue arrows mark a microscopic line structure, both of which are commonly seen in graphene films fabricated with CVD methods. Scale bar, 10 μm. **b,** A typical Raman spectrum of our graphene film away from any dark spots or dark lines shown in **a**. **c**, Scattering amplitude $s(\omega=1150\ cm^{-1})$ at various gate voltages $V_g$ normalized to that at $V_g$=0V.

Raman spectroscopy (Senterra, Bruker Inc.) was applied to characterize our graphene films. All our Raman measurements were carried out using a 532 nm excitation laser, a 50× (NA=0.75) objective, and a grating with 1200 lines per millimeter. The laser spot size is roughly 1 μm, and the spectral resolution is 3 cm$^{-1}$. An accuracy of ~1 cm$^{-1}$ can be achieved by band-fit when determining the peak positions for G and 2D bands. We kept our laser power below 2 mW to avoid heating[2]. Raman spectra were collected all across our graphene films to characterize our film quality and doping level. A typical spectrum taken away from any dark spots or dark lines (Fig. S1a) is given in Fig. S1B. A symmetric 2D peak verified that our film is a single layer graphene, while a vanishing D peak indicates that our film is of high crystalline quality. According to previous studies, the G peak position is sensitive to the doping level of graphene[3-5]. The average G peak position of Raman spectra taken at different locations is around 1595±1 cm$^{-1}$ indicating extremely high doping in our CVD graphene film.

To estimate the carrier polarity and density of our graphene film, we investigated the gating dependence of the near-field IR response by monitoring the hybrid plasmon-phonon resonance around $\omega$=1150cm$^{-1}$. At this frequency, the scattering amplitude $s$ scales monotonically with the doping level of graphene (see Ref. S8 for detailed information), thus offering a convenient way to estimate the doping level of graphene. As shown in Fig. S1c, $s(\omega=1150\ cm^{-1})$ decreases systematically with increasing gate voltage $V_g$. The charge neutral point $V_{CN}$ is above $V_g = 80$ V and exceeds the breakdown voltage of the SiO$_2$ layer in our structure. Albeit incomplete, these gating results nevertheless conclusively show that our graphene films are highly hole-doped at ambient conditions.

Based on the combination of our Raman and near-field gating experiments, we estimated that the hole density of our CVD graphene film was around $(1.0\pm0.3)\times10^{13}$ cm$^{-2}$. The corresponding Fermi energy $E_F$ is about $0.37\pm0.06$ eV estimated from $E_F = \hbar v_F \sqrt{\pi n}$, where $v_F \approx 1\times10^6$ m/s is the Fermi velocity. This high level of doping likely originates from both SiO$_2$ substrate and molecule adsorption in air atmosphere[6,7].

## 2. Nomenclature of line defects

In addition to the cracks and grain boundaries (GBs) introduced in the main text, we also found other types of line defects including wrinkles and grain-overlaps. In Fig. S2, we plot both atomic force microscopy (AFM) (Figs. S2a and S2d) and scanning plasmon interferometry (SPI) (Figs. S2b and S2e) images for these two types of line defects. All SPI images were taken at $\lambda_{IR}$=11.3 μm and share the same color scale. For the purpose of quantitative analysis, in the right panels of Fig. S2, we plot the line profiles across the twin fringes of these line defects.

Wrinkles (i.e. film corrugations) in CVD graphene are formed during either post-growth cooling or film transfer processes[1]. Here we only discuss wrinkles in the nanometer length scale. As shown in Fig. S2b, wrinkles also generate twin fringes similar to cracks and GBs indicating that they also reflect surface plasmons (SPs). Nevertheless, the fringe intensity and separation between the twin fringes $D_{TF}$ for wrinkles are different from position to position (Figs. S2b). Such differences are due to the variations of the structural morphologies[9] of these wrinkles at different locations.

Grain-overlaps are line defects formed when one grain overlaps with another, so that they bridge different grains[10]. Unlike GBs, grain overlaps are clearly visible in AFM topography. There are two grain-overlaps here in Fig. S2d (marked with OL1 and OL2), producing only ~1 nm variation in the AFM topography. Despite their similarity in the topography, OL1 and OL2 trigger totally different twin fringes (Fig. S2e). The twin fringes of OL1 are very close to each other, while those of the OL2 are much further apart. Both of them are different from the twin fringes triggered by a GB (marked with a red arrow in Fig. S2e). The different SPI response of the two grain-overlaps might be related to the stacking order of the overlapped region.

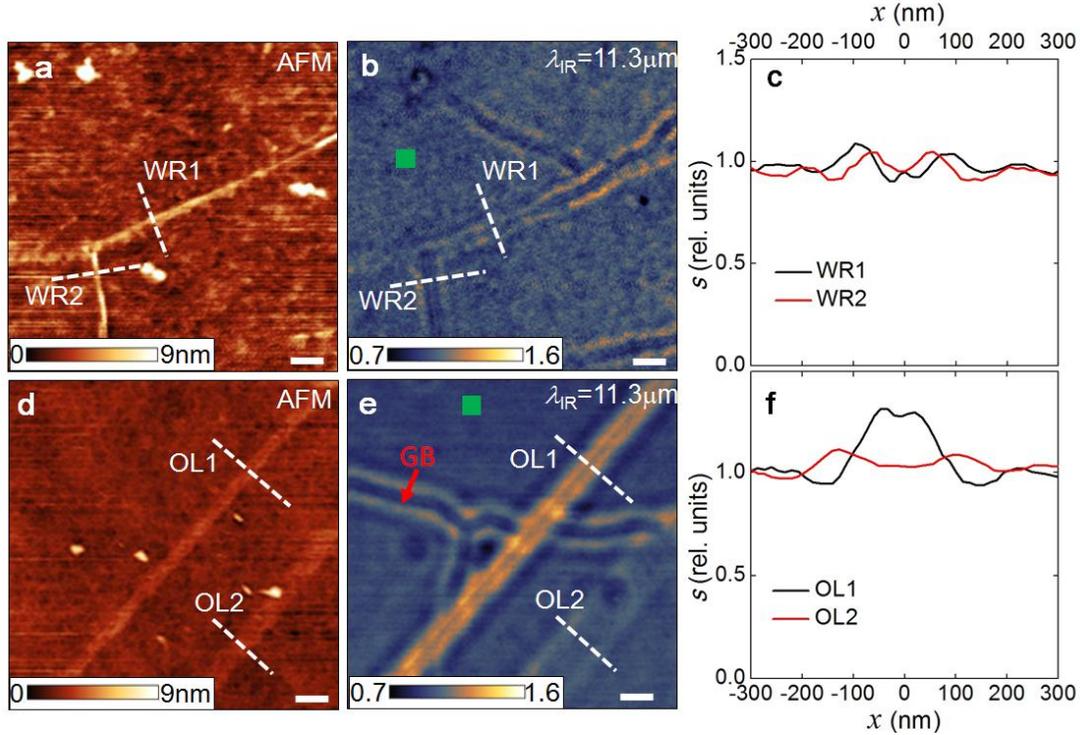

**Figure S2 | Wrinkles and grain-overlaps. a.** Topography image showing wrinkles. **b.** SPI image taken simultaneously with **a** at $\lambda_{IR}$=11.3 μm. WR1 and WR2 in **a** and **b** mark the two wrinkles. **c.** The line profiles taken along the dashed lines in **b**. **d.** Topography image of grain-overlaps. **e.** SPI image taken simultaneously with **d** at $\lambda_{IR}$=11.3 μm. Red arrow marks a GB. OL1 and OL2 in **c** and **d** mark two different types of grain-overlaps. **f.** The line profiles taken along the dashed lines in **e**. In both **c** and **f**, the scattering amplitude *s* is normalized to the places far away from the line defects where no plasmons fringes exist (e.g. green squares in **b** and **f**). Scale bars in all panels are 200 nm.

In Fig. S3, we show a larger-area scan of our CVD graphene film including various types of line defects. Based on the AFM topography (Fig. S3a) and SPI (Fig. S3b) images, we were able to sketch a map for various types of line defects (Fig. S3c). Topographic and SPI signatures allowing us to distinguish different types of line defects are described in the manuscript and the above paragraphs. Being sub-nm wide defects, GBs have no obvious topography features, yet they trigger clearly observable plasmonic twin fringes. Grain overlaps and wrinkles show up in both the AFM topography and the SPI images. The main difference between grain-overlaps and wrinkles is the degree of continuity and the intensity of the twin fringes. The wrinkles are sporadic and discontinuous with fringe intensity varying from position to position. The grain-overlaps are continuous (similar to GBs) with almost constant fringe intensities. High-resolution AFM and SPI images (like Figs. S1 & S2) are well suited to discriminate between all these different types of line defects.

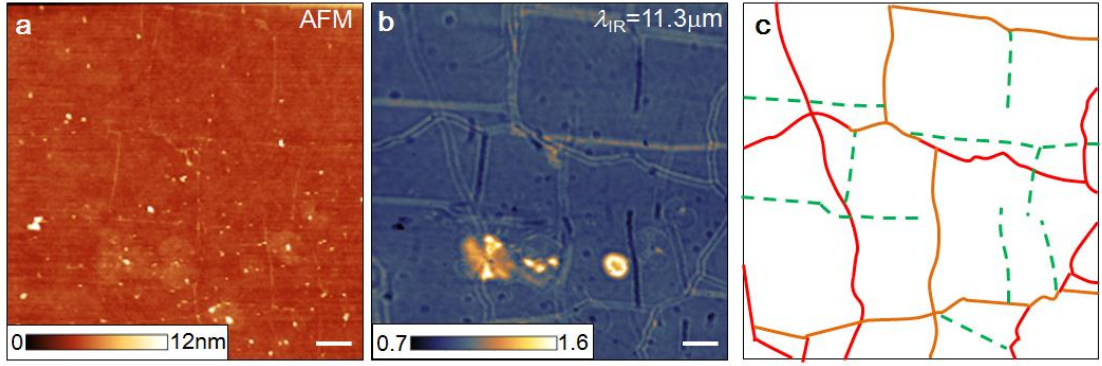

**Figure S3 | Large-area scanning revealing various types of line defects. a**. AFM topography image. **b**. SPI image simultaneously taken with **a** at $\lambda_{IR}$=11.3 μm. **c**. The map of various types of line defects including GBs (red), grain-overlaps (orange), and wrinkles (green). Scale bar width in all the panels is 1 μm.

### 3. Reflection of plasmons from a linear defect

The observed fringes originate from interference of the plasmon waves launched by the AFM tip and those backscattered by a linear defect. Here we only consider line defects with negligible geometric width, such as a GB. Theoretical modeling of such waves is a challenging problem that requires solving complicated integro-differential equations. The problem becomes more manageable once one introduces certain approximations for the response functions of graphene and the tip, as described in our previous work[11]. However, even after these approximations the solution can be obtained only numerically. Before we go into details of our numerical simulations (Section 5), we first consider a simpler scattering problem, which can be tackled analytically.

Instead of a complicated waveform launched by the tip, we consider a plane wave incident from the left on the line defect located at $x = 0$. We take the scalar potential of this wave to be $\phi(x,y) = e^{iq_x x + i q_y y}$ in the graphene plane. The system is assumed to be uniform along $y$, so that $q_y$ is conserved. The $x$-component of the incident plasmon momentum is

$$q_x = \sqrt{q_\infty^2 - q_y^2}, \ \mathrm{Im}\, q_x \geq 0, \qquad (S1)$$

where $q_\infty \equiv q_p(\infty)$, function $q_p(x)$ is the local plasmon momentum,

$$q_p(x) = \frac{i\kappa\omega}{2\pi\sigma(x)}, \ \kappa = \frac{\varepsilon_{sub}+1}{2}, \qquad (S2)$$

$\sigma(x)$ is the local sheet conductivity of graphene, and $\kappa$ is the effective dielectric constant. We parameterize the deviation of $q_p(x)$ from its limiting value at infinity by the dimensionless function $g(x)$ such that

$$\frac{1}{q_p(x)} = \frac{1+g(x)}{q_\infty}. \qquad (S3)$$

We assume that $g(x)$ rapidly decays with $x$ (faster than $1/x$). Note that the plasmon wavelength discussed in the main text is defined by $\lambda_p(x) \equiv 2\pi / q_1$ with $q_1 = \mathrm{Re}\, q_p(x)$

Our goal is to calculate the potential $\psi(x)e^{iq_y y}$ of the scattered wave. In particular, we are interested in the behavior of $\psi(x)$ at large negative $x$,

$$\psi(x) \approx r_{sp} e^{-iq_x x} = |r_{sp}| e^{-iq_x x + i\delta_{sp}}, \tag{S4}$$

which defines the reflection probability $|r_{sp}|^2$ and the phase shift $\delta_{sp}$ of graphene plasmons.

Our starting equations are:

$$\Phi(x) \equiv \phi(x,0) + \psi(x), \tag{S5}$$

$$\Phi(x) \equiv V_1 * \{\frac{1+g(x)}{q_\infty} q_y^2 \Phi(x) - \partial_x \frac{1+g(x)}{q_\infty} \partial_x \Phi(x)\}, \tag{S6}$$

where $\Phi(x)$ is the total potential, $V_1 = \pi^{-1} K_0(q_y x)$ is the 1D Fourier transform of the Coulomb potential, $K_0(z)$ is the modified Bessel function of the second kind, and the star denotes convolution,

$$\Phi(x) \equiv \phi(x,0) + \psi(x), \tag{S7}$$

We approach Eq. (S6) using the Green's function perturbation theory method. The Green's function is defined by

$$G(x) = \int \frac{dk'}{2\pi} \frac{e^{ikx}}{\varepsilon(k,q_y)}, \quad \varepsilon(k_x, k_y) = 1 - \frac{\sqrt{k_x^2 + k_y^2}}{q_\infty}. \tag{S8}$$

The physical meaning of $G$ is the response to the localized disturbance; $\varepsilon(k_x, k_y)$ is the 2D dielectric function of graphene. Using contour-integration techniques, the Green's function can be split into two terms:

$$G(x) = -\frac{iq_\infty^2}{q_x} e^{iq_x |x|} + \Delta G(x). \tag{S9}$$

The first term represents the outgoing plane wave and the second term is a correction decaying as $\Delta G(x) \sim (q_y x)^{-3/2}$ for $q_y \neq 0$ and $\Delta G(x) \sim 2/(q_\infty x)^2$ for $q_y = 0$. In the latter case, $\Delta G(x)$ can be expressed in terms of the standard special functions, the cosine-integral Ci($z$) and the sine-integral Si($z$):

$$\Delta G(x) = -\frac{q_\infty}{\pi} \{\text{Ci}(q_\infty x) \cos q_\infty x + [\text{Si}(q_\infty x) - \frac{\pi}{2}] \sin q_\infty x\}. \tag{S10}$$

Using thus defined Green's function, Eq. (S6) can be transformed to

$$\psi(x) = \frac{1}{q_\infty} (G*V_1) * \{g(x) q_y^2 \Phi(x) - \partial_x g(x) \partial_x \Phi(x)\}, \tag{S11}$$

which is analogous to the Lippmann-Schwinger equation of the usual scattering theory. Following the familiar route, at $x$ much longer than plasmon wavelength $\lambda_p$, we neglect the correction $\Delta G(x)$ in $G(x)$ and recover Eq. (S5) with the following reflection coefficient:

$$r_{sp} = -\frac{i}{q_x} \int_{-\infty}^{\infty} dx e^{iq_x x} [q_y^2 g(x) \Phi(x) - \partial_x g(x) \partial_x \Phi(x)]. \tag{S12}$$

We restrict our further analysis to the case of a weak defect, i.e., small $g(x)$. In this case $|r_{sp}| \ll 1$, $\Phi(x) \approx e^{iq_x x}$, and the formula similar to the first Born approximation applies:

$$r_{sp} \approx i\frac{q_x^2 - q_y^2}{q_x} \tilde{g}(-2q_x), \quad \tilde{g}(k) \equiv \int_{-\infty}^{\infty} dx\, e^{-ikx} g(x). \tag{S13}$$

Notably, the reflection vanishes at the "Brewster angle" of $\pi/4$ where $q_x = q_y$. However, we are primarily interested in the normal incidence ($q_y = 0$). The most important for us is the situation where the effective electronic width of the defect is small compared to the plasmon wavelength: $W_{eff} \ll \lambda_p$. In this case, for $q_y = 0$, Eq. (S13) acquires a remarkably simple form

$$r_{sp} \approx iW_{eff}\Delta q_p, \quad \Delta q_p \equiv \frac{1}{W_{eff}} \int [q_p(x) - q_\infty] dx. \tag{S14}$$

Parameter $\Delta q_p$ has the meaning of the average deviation of $q_p(x)$ inside the defect region from its limiting value $q_\infty$. In turn, the phase shift of graphene plasmons is given by

$$\delta_{sp} = \arg(iW_{eff}\Delta q_p). \tag{S15}$$

For real $\Delta q_p$, $\delta$ can take only two values: $\delta_{sp} = -\pi/2$ if $\Delta q_p < 0$, (i.e., $\lambda_p$ inside the defect is higher than outside), and $\delta_{sp} = -3\pi/2$ otherwise. On the other hand, if $\Delta q_p$ also has an imaginary part, the phase shift can be arbitrary.

## 4. Understanding the interference patterns

Let us now apply the above results to the task of interpreting the positions of the interference fringes found in the experiment, i.e., the tip positions $\boldsymbol{\rho}_t = (x_t, y_t)$ where the nanoscope registers the maxima of the signal $s(\boldsymbol{\rho}_t)$. The relation between $s(\boldsymbol{\rho}_t)$ and the previously discussed scalar potential $\Phi(\boldsymbol{\rho})$ is complicated and in fact tip-dependent[8]. However, according to our numerical simulations, the maxima of $s(\boldsymbol{\rho}_t)$ occur roughly where the scalar potentials $\phi(\boldsymbol{\rho})$ and $\psi(\boldsymbol{\rho})$ due to, respectively, the launched and the scattered waves, add in phase underneath the tip. The results of the previous section can be straightforwardly utilized provided the tip is located far away from the linear defect. Assuming that is the case, let us discuss the launched wave $\phi(\boldsymbol{\rho})$ first. Near the defect, which is far from the tip, $\phi(\boldsymbol{\rho})$ behaves as an outgoing cylindrical wave:

$$\phi(\boldsymbol{\rho}) \approx C_0 e^{i\delta_t} \frac{\phi(\boldsymbol{\rho}_t)}{\sqrt{q_p |\boldsymbol{\rho} - \boldsymbol{\rho}_t|}} e^{-iq_p |\boldsymbol{\rho} - \boldsymbol{\rho}_t|}, \quad q_p |\boldsymbol{\rho} - \boldsymbol{\rho}_t| \gg 1. \tag{S16}$$

The coefficient $C_0 \sim 1$ and the phase shift $\delta_t$ depend on microscopic parameters of the tip, graphene, and the substrate. There is no general reason for $\delta_t$ to be negligible.

Next, consider the reflected wave $\psi(\boldsymbol{\rho})$. To compute this function, one can decompose $\phi(\boldsymbol{\rho})$ into Fourier harmonics with all possible $q_y$, determine the reflected wave for each harmonic, and then evaluate the inverse Fourier transform at the tip position. It is easy to see that the reflected wave is dominated by harmonics of nearly normal incidence, $q_y \sim |q_p/x_t|^{1/2} \ll q_p$. This allows one to replace function $r(q_y)$ in this calculation by the constant $r(0)$. In turn, it means that the method of images applies, so that $\psi(\boldsymbol{\rho})$ can be approximated by a cylindrical wave of a certain amplitude radiated from the position $(-x_t, y_t)$. This argument is the theoretical basis for the illustration shown in Fig. 1A of the main text. Adding together the launched and the reflected waves, we find the total potential at the tip position:

$$\frac{\Phi(\boldsymbol{\rho}_t)}{\phi(\boldsymbol{\rho}_t)} \approx 1 + \frac{C_0 |r(0)|}{\sqrt{2q_p |x_t|}} e^{2iq_p|x_t|+i\delta_g+i\delta_t}, \quad q_p |x_t| \gg 1. \tag{S17}$$

According to the earlier assumption, the interference maxima occur when $2q_1|x_t|+\delta_{sp}+\delta_t$ is an integer multiple of $2\pi$. They form a sequence of equidistant points on each side of the defect:

$$|x_t| = \frac{\lambda_p}{2}\left(n + \{-\frac{\delta_{sp}+\delta_t}{2\pi}\}\right), \tag{S18}$$

where $n = 0, 1, \ldots$ and $\{z\}$ stands for the fractional part of $z$. Although Eq. (S18) was derived assuming $n \gg 1$, it should not be grossly incorrect at $n = 0$. Therefore, the separation between the maxima nearest to the defect is:

$$D_{TF} \approx \left\{-\frac{\delta_{sp}+\delta_t}{2\pi}\right\}\lambda_p. \tag{S19}$$

Thus the magnitude of $D_{TF}$ is governed both by the plasmon phase shift $\delta_{sp}$, and by the tip-dependent parameter $\delta_t$. Based on our numerical modeling results given in Fig. S5 and Eqs. S15 & S19, we were able to estimate $\delta_t$ to be $-(0.5\pm0.1)\pi$, which is fairly robust for tip radius from 10 nm to 100 nm. The estimation was done by comparing $D_{TF}$ inferred using Eq. S19 with that obtained from modeled profiles. The value of $-(0.5\pm0.1)\pi$ is fairly accurate for all the test modeling profiles. Slight deviation (less than 20%) occurs when $g(x)$ or $\Delta q_p$ is relatively big, i.e. when $\lambda_p^{LD}$ is 100 nm or 800 nm in Fig. S5a.

## 5. Numerical modeling of twin fringes profiles

Many elements of our numerical modeling have already been described in ref. S11. In short, we model our AFM tip as a metallic spheroid (Fig. S4a): the length of the spheroid is $2L$ and the radius of curvature at the tip end is $R$. Here, $R$ is set to be 25 nm according to the manufacturer specification and $L$ is not a sensitive parameter so long as it is much larger than $R$ ($L$ is set to be $9R$ in all our simulations). The scattering amplitude $S$ (before demodulation) is proportional to the total radiating dipole $p_z$ of the spheroid. Therefore, in order to fit the line profiles perpendicular to the twin fringes due to a line defect, we need to calculate $p_z$ at different spatial coordinates $(x, z_{tip})$. Here, $x$ and $z$ are the $x$- and $z$- coordinates of lower end of the AFM tip, respectively. In order to compute $p_z$, we assume that the electric potential $\Phi$ outside both the tip and the sample can be represented as a superposition of potentials of a large number of point dipoles positioned inside the tip. Based on this assumption, we are able to calculate the electric potential $\Phi$ and field $\mathbf{E}$ distribution at every given point of the space. Imposing the boundary condition that the component of $\mathbf{E}$ tangential to the tip is zero, we obtain individual dipole moments. The total dipole moment $p_z$ of the tip is their sum. By calculating $p_z$ at different $z$, we are able to perform 'demodulation' of the scattering amplitude $S$ and get different harmonics of the scattering signal. While calculating $p_z$ at different $x$ allows us to plot the modeling scattering amplitude and phase profiles. In all our modeling and simulation, we assume no position dependence in the $y$- direction for the purpose of simplicity. In the current work, the scattering amplitude $s$ is normalized to that far away from the line defect where no plasmon fringes exist.

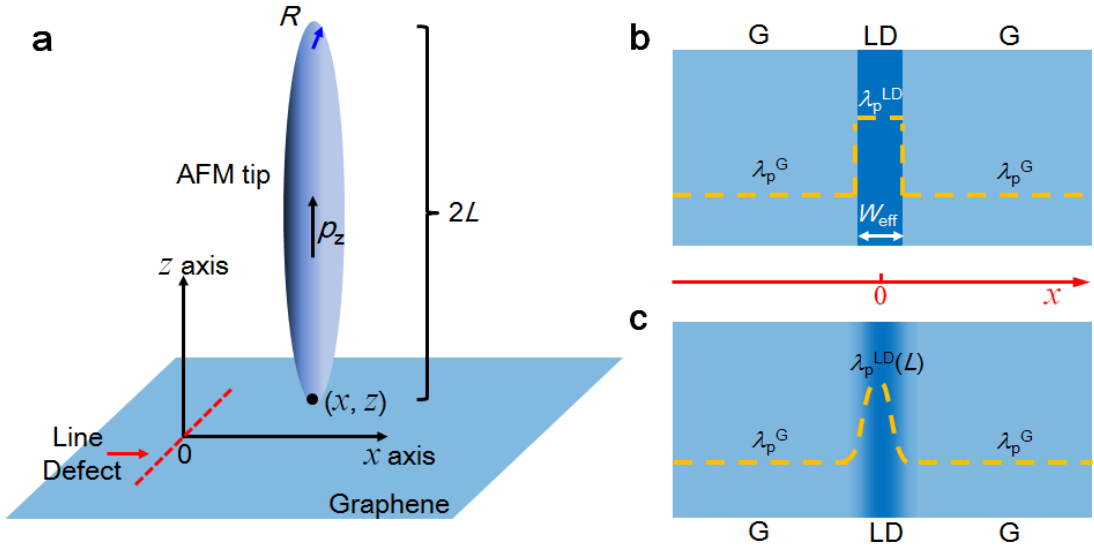

**Figure S4 | Modeling of the AFM tip and graphene. a.** Modeling parameters of the AFM tip. Red dashed line marks a line defect of graphene at $x=0$. **b.** 'Discontinuous' model for a line defect in graphene. **c.** 'Gradual' model for a line defect in graphene. 'G' and 'LD' in both **b** and **c** stand for graphene film and line defect, respectively. The fill colors and the yellow dashed lines in both **b** and **c** illustrate the variation of modeling parameters such as $\lambda_p^G$ (as plotted) or $\gamma_p^G$.

5.1 Model with discontinuous change of parameters

As for graphene, in our previous work[11] we used the complex plasmon wavevector $q_p$ (Eq. S2) as an input parameter in our modeling. Equivalently here, the plasmon wavelength $\lambda_p = 2\pi/\mathrm{Re}(q_p)$ and the damping rate $\gamma_p = \mathrm{Im}(q_p)/\mathrm{Re}(q_p)$ are the input parameters. We start with a model that assumes that graphene has a constant plasmon wavelength $\lambda_p^G$ and damping rate $\gamma_p^G$ away from the line defect, and that a line defect with an effective width of $W_{\mathrm{eff}}$ is characterized by its own plasmon wavelength $\lambda_p^{LD}$ and damping rate $\gamma_p^{LD}$ as illustrated in Fig. S4b. Here and below, this model is referred to as the Discontinuous Model. Among the four parameters, both $\lambda_p^G$ and $\gamma_p^G$ can be estimated directly from our experimental data. $\lambda_p^G$ is set to be around 260 nm by measuring the fringe width of two side fringes at $|x|\approx 230$ nm. $\gamma_p^G$ is estimated to be around 0.15 by comparing the plasmon damping to that of exfoliated graphene[11].

To understand how $\lambda_p^{LD}$, $\gamma_p^{LD}$ and $W_{\mathrm{eff}}$ affect the plasmon fringe profile, we first perform a series of modeling by varying only one parameter and fixing the other two constant. In Fig. S5, we show fours representative sets of modeling results by:
(1) varying $\lambda_p^{LD}$ from 10 to 800 nm with $\gamma_p^{LD}=\gamma_p^G=0.15$ and $W_{\mathrm{eff}}\approx 30$ nm (Fig. S5a);
(2) varying $\gamma_p^{LD}$ from 0.01 to 2.0 with $\lambda_p^{LD}=500$ nm and $W_{\mathrm{eff}}\approx 30$ nm (Fig. S5b);
(3) varying $\gamma_p^{LD}$ from 0.01 to 2.0 with $\lambda_p^{LD}=100$ nm and $W_{\mathrm{eff}}\approx 30$ nm (Fig. S5b);
(4) varying $W_{\mathrm{eff}}$ from 5 to 80 nm with $\lambda_p^{LD}=500$ nm and $\gamma_p^{LD}=0.5$ (Fig. S5d).

In all panels of Fig. S5, we plot the modeling scattering amplitude $s$ profiles along with the experimental data for a GB in Fig. 2 in the main text. The scattering amplitude $s$ is normalized to its value far away from the line defect, $|x| \geq 300$ nm in Fig. S5. We monitor the evolution of both the fringe intensity (peak height) and the separation between twin fringes $D_{TF}$ with varying $\lambda_p^{LD}$, $\gamma_p^{LD}$ or $W_{eff}$. As explained above, the fringe intensity is related to the reflection probability $|r|^2$ (Eq. S14), while $D_{TF}$ is determined by the phase shift $\delta_{sp}$ (Eq. S19).

As one can see in Fig. S5a, the further $\lambda_p^{LD}$ deviates from $\lambda_p^G$, the higher the fringe intensity is. This is consistent with Eq. S14 since larger $|\lambda_p^{LD} - \lambda_p^G|$ leads to larger $\Delta q_p$ and hence higher reflection probability $|r|^2$. The separation between the twin fringes $D_{TF}$ also depends on $\lambda_p^{LD}$. Assuming $\lambda_p^{LD} > \lambda_p^G$, we can find simulation parameters that bring $D_{TF}$ close to the experimentally observed width 150 nm. Conversely, if we assume that $\lambda_p^{LD} < \lambda_p^G$, the magnitude of $D_{TF}$ becomes too large, about 260nm, nearly twice the observed value. This is again consistent with the analytical theory above. When $\lambda_p^{LD} - \lambda_p^G$ switches its sign, plasmon phase shift $\delta_{sp} = \arg(iW_{eff}\Delta q_p)$ will be shifted by $\pi$ (Eq. S15), resulting in drastic change in $D_{TF}$ (Eq. S19).

Now we examine the effects of $\gamma_p^{LD}$ on the plasmon fringe profiles. In Figs. S5b and S5c, we show the modeling results with $\lambda_p^{LD}$ fixed at 500 nm and 100 nm, respectively. In both cases, the fringe profile evolves systematically with varying $\gamma_p^{LD}$, in agreement with Eqs. S14 & S19. Notably, in the case of $\lambda_p^{LD} = 500$ nm, the scattering amplitude $s$ at the line defect ($x \approx 0$) shows a sensitive dependence on $\gamma_p^{LD}$. Good agreement with the experimental data can be achieved only if $\lambda_p^{LD} > \lambda_p^G$, i.e., if the GB is more doped than the rest of the film

The modeling results for several $W_{eff}$ are presented in Fig. 4d, where one can see that the fringe intensity decreases rapidly with decreasing $W_{eff}$. This is because $|r|$ scales with $W_{eff}$ as shown in Eq. S14. At the smallest $W_{eff} \approx 5$ nm, the twin fringes almost disappear. As $W_{eff}$ increases, the separation between the twin fringes $D_{TF}$ increases by about the same amount.

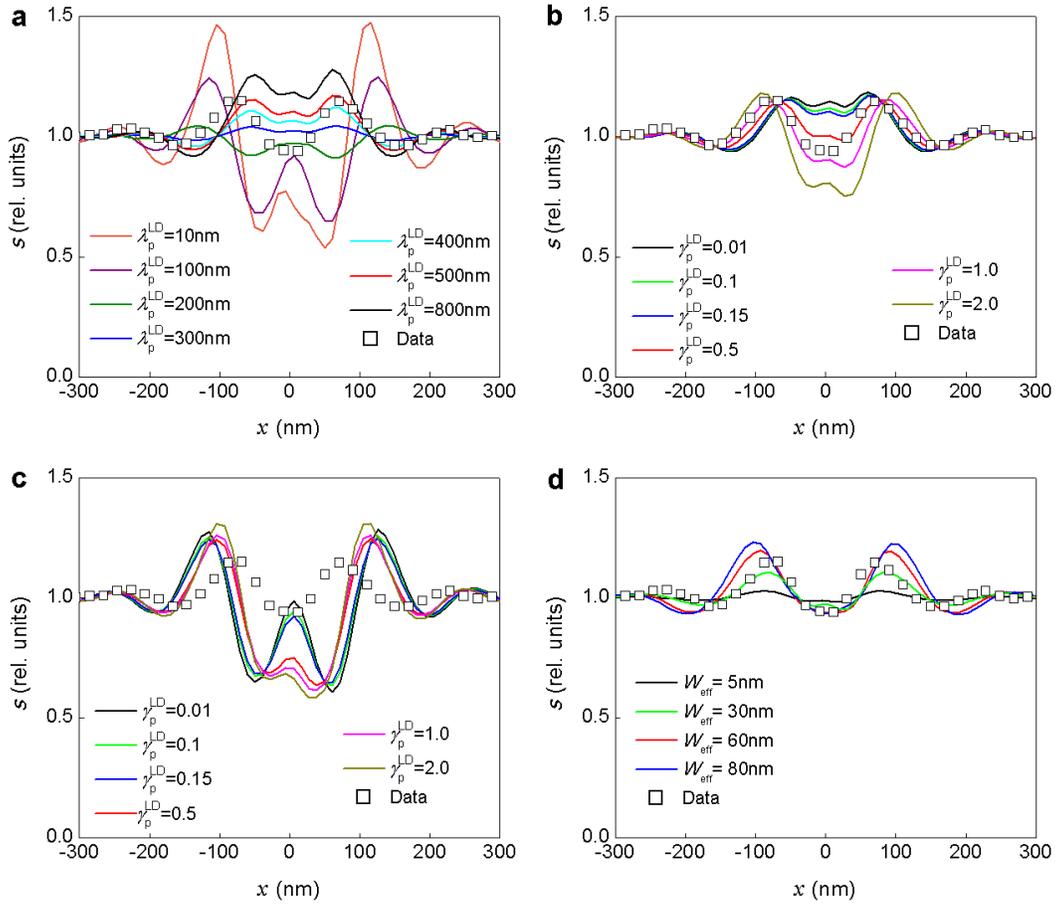

**Figure S5 | Fringe profile simulation with the Discontinuous model. a.** Modeling $s$ profiles with 10 nm$<\lambda_p^{LD}<$800 nm, $\gamma_p^{LD}=\gamma_p^{G}=0.15$ and $W_{eff}\approx$30 nm. **b.** Modeling $s$ profiles with 0.01$<\gamma_p^{LD}<$ 2, $\lambda_p^{LD}$ =500 nm and $W_{eff}\approx$30 nm. **c.** Modeling $s$ profiles with 0.01$<\gamma_p^{LD}<$ 2, $\lambda_p^{LD}$ =100 nm and $W_{eff}\approx$30 nm. d. Modeling $s$ profiles with 5nm$<W_{eff}<$ 90 nm, $\lambda_p^{LD}$ =500 nm and $\gamma_p^{LD}$ =1.0. In all panels, $\lambda_p^{G}$=260, $\gamma_p^{G}$=0.15, the line defect is at $x$=0, and experimental data of a GB taken at 11.26 μm is plotted with black hollow squares. Slightly asymmetry in our modeling results is due to limited resolution of our modeling.

Figure S5d illustrates how the calculated $s(x)$ profiles change as a function of a single parameter of the set ($\lambda_p^{LD}$, $\gamma_p^{LD}$ or $W_{eff}$) while the remaining ones are kept fixed. Finally, in Fig. S6, we vary all the three parameters in order to get the best fit to the data. Such a fit is achieved with $W_{eff}$ close to 20 nm, which is much larger than its geometric width < 1 nm. Effective widths much smaller than 20 nm, e.g., $W_{eff}\approx$5 nm，require setting $\lambda_p^{LD}$ as high as 3000 nm to fit the data, corresponding to an unrealistic carrier density of $n$=1.2×$10^{15}$ cm$^{-2}$.

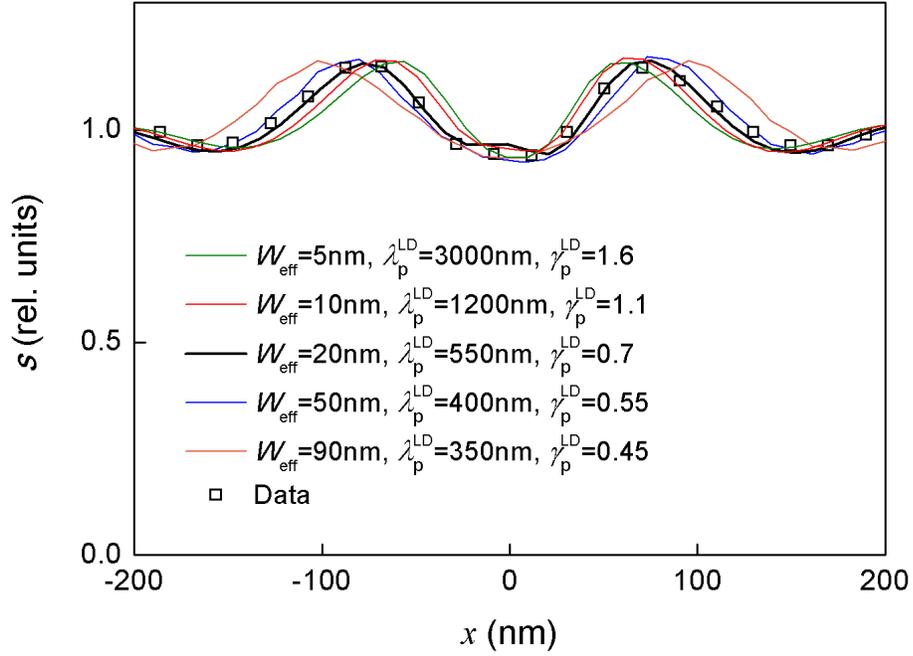

**Figure S6 | Fringe profile fitting with the Discontinuous model.** Calculated $s(x)$ profiles for five different effective widths $W_{eff}$ =5, 10, 20, 50, 90 nm. For each $W_{eff}$, $\lambda_p^G$=260, $\gamma_p^G$=0.15 are fixed but $\lambda_p^{LD}$ and $\gamma_p^{LD}$ are adjusted to best reproduce the experimental data (squares) taken at 11.26μm.

5.2 Model with a gradual change of parameters

So far, for the purpose of simplicity and clarity, we use the Discontinuous model (Fig. S2B) for calculation. Clearly, the model grasps the gross features of the experimental data. Nevertheless, in this model both $\lambda_p(x)$ and $\gamma_p(x)$ profiles have discontinuities close to the line defect (Figs. S5). We also considered a more realistic model that was referred to as the Gradual Model (Fig. S2B). In this model the rapid increase of both $\lambda_p$ and $\gamma_p$ close to the line defect is modeled by exponential functions:

$$\lambda_p(x) = \lambda_p^G + A_1 e^{-2|x|/B}, \quad \gamma_p(x) = \gamma_p^G + A_2 e^{-2|x|/B}. \tag{S20}$$

Here $A_1$, $A_2$ and $B$ are the new adjustable parameters. $A_1$ and $A_2$ determine the peak height of $\lambda_p$ and $\gamma_p$ at the center of the line defect, respectively (Fig. S2C), $B$ is associated with the effective width of the line defect.

In Fig. S7, we show the best-fit results for the GB data taken at $\lambda_{IR}$ from 10.7 to 11.3 μm using the Gradual model. The modeling parameters are: $A_1$=320 nm, $A_2$=0.9, $B$=20 nm, corresponding $\lambda_p(x)$ and $\gamma_p(x)$ profiles are plotted in Fig. 2e in the main text.

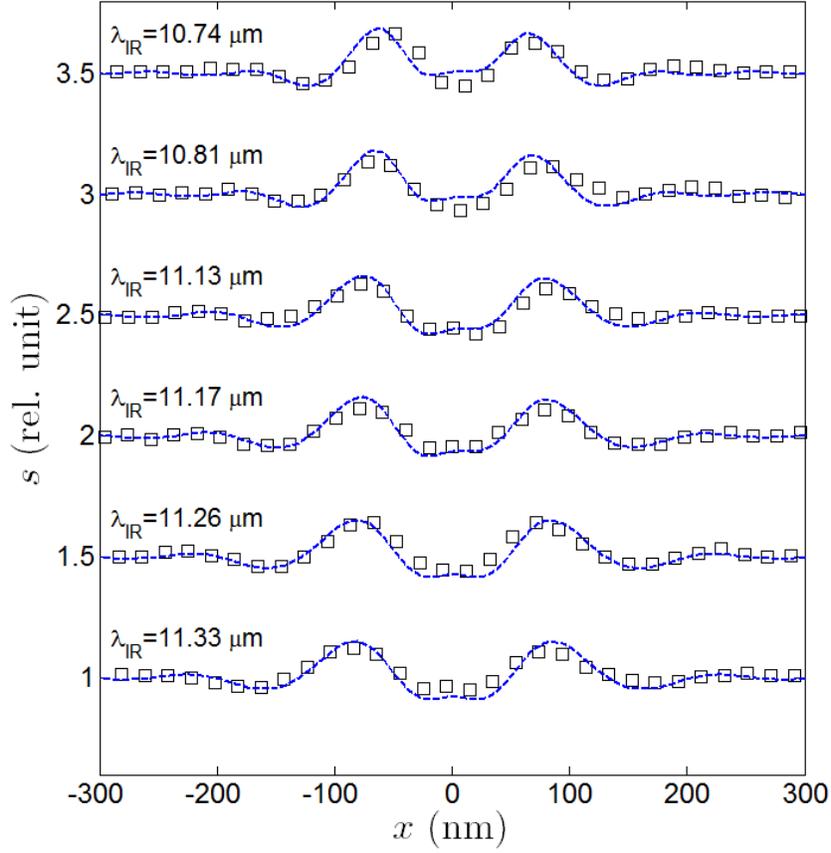

**Figure S7 | Fitting of twin fringe profiles with the Gradual model. a.** Line profiles across twin fringes at various $\lambda_{IR}$ obtained from both experimental data of a GB (black squares) and modeling (blue dashed curve). Here, the scattering amplitude $s$ is normalized to that far away from the line defect ($|x|>300$ nm). All line profiles are vertically displaced for clarity.

### 6. Discussion

Our modeling with both models not only fits well the experimental data, but also uncovers many essential properties of GBs. (1) GBs tend to have higher $\lambda_p$ compared to the rest of CVD film. (2) GBs tend to have higher $\gamma_p$ compared to the rest of CVD film. (3) GBs tend to have higher effective width ($W_{eff}$~20 nm) than their geometric width ($W<$ 1nm).

6.1 The effective electronic width of grain boundaries

According to Eq. 1 in the main text, plasmon wavelength $\lambda_p$ of graphene is proportional to its Fermi energy $E_F$ (Eq. 1 in the main text). Considering that $E_F = \hbar v_F \sqrt{\pi n}$ ($v_F$ is the Fermi velocity, $n$ is the carrier density of graphene), higher $\lambda_p$ implies an increase of the carrier densities in the vicinity of GBs. This is expected since GBs are lattice imperfections that favor molecule adsorptions at ambient conditions[12,13], which will further enhance the hole doping in ambient[14,15]. Within the Drude approximation, the plasmon damping rate $\gamma_p$ in graphene can be written as[11]

$$\gamma_p = \frac{q_2}{q_1} \approx \frac{\kappa_2}{\kappa_1} + \frac{\sigma_1}{\sigma_2} = 0.05 + \frac{1}{\omega\tau}, \tag{S21}$$

here $\kappa_1$ and $\kappa_2$ are the real and imaginary parts of effective dielectric constant $\kappa$ (Eq. 2), and $\sigma_1$ and $\sigma_2$ are real and imaginary parts of optical conductivity of graphene $\sigma$, $1/\tau$ is the scattering rate of the charge carriers (as labeled in Fig. S6c). Therefore, higher $\gamma_p$ indicates higher scattering rate $1/\tau$ close to the GBs. These additional scattering originates presumably from the strong structural and Coulomb disorder at the GB.

The effective electronic width $W_{\text{eff}} \sim 20$ nm of the GBs revealed by the SPI is comparable to the screening length in graphene and is much larger than the sub-nm geometric width of the grain boundaries. The relevant screening problem has been considered in ref. S16 within the perturbation-theory approach. A non-perturbative treatment in ref. S17 yields qualitatively similar results apart from logarithmic corrections.

6.2 Charge transport and plasmon propagation at grain boundaries

Starting from the $E_F$ and $1/\tau$ profiles displayed in Figs. S6b and S6c, we are able to calculate the DC conductivity profile across the GB using the formula obtained under Drude approximation:

$$\sigma_{DC} = \frac{2e^2}{h} \frac{E_F}{E_\Gamma}, \tag{S22}$$

here $E_\Gamma = \hbar\tau^{-1}$ is the scattering energy. The obtained $\sigma_{DC}(x)$ is given in the inset of Fig. 2c of the main text, where one can see that the GB tends to have a lower DC conductivity compared to the rest of the graphene film.

Previous transport and STM studies[18-21] of GBs were all performed in vacuum. Graphene was much less doped in those studies. On the contrary, our experiments were carried out in ambient atmospheric conditions, thus revealing for the first time the transport properties of GBs in graphene films that are highly hole-doped (presumably, by oxygen and water molecules). We remark that the 'electronic' nature of the GBs are the origin for the lower DC conductivities observed in our experiments.

**References**

1. Li, X. et al. Large-area synthesis of high-quality and uniform graphene films on copper foil. Science 324, 1312-1314 (2009).

2. Ferrari, A. C. et al. Raman spectrum of graphene and graphene layers. Phys. Rev. Lett. 97, 187401 (2006).

3. Pisana, S. et al. Breakdown of the adiabatic Born-Oppenheimer approximation in graphene. Nature Mater. 6, 198-201 (2007).

4. Das, A. et al. Monitoring dopants by Raman scattering in an electrochemically top-gated graphene transistor. Nature Nanotech. 3, 210-215 (2008).

5. Kalbac, M. et al. The influence and strong electron and hole doping on the Raman intensity of chemical vapor-deposition graphene. ACS Nano 4, 6055-6063 (2010).

6. Ryu, S. et al. Atmospheric oxygen binding and hole doping in deformed graphene on a SiO2 substrate. Nano Lett. 10, 4944-4951 (2010).

7. Sojoudi, H., Baltazar, J., Henderson, C. & Graham, S. Impact of post-growth thermal annealing and environmental exposure on the unintentional doping of CVD graphene films. J. Vac. Sci. Technol. B 30, 041213 (2012).

8. Fei, Z. et al. Infrared Nanoscopy of Dirac Plasmons at the Graphene-SiO2 interface. Nano Lett. 11, 4701-4705 (2011).

9. Zhu, W. et al. Structure and electronic transport in graphene wrinkles. Nano Lett. 12, 3431-3436 (2012)

10. Tsen, A. W. et al. Tailoring electrical transport across grain boundaries in polycrystalline graphene. Science 336, 1143-1146 (2012).

11. Fei, Z. et al. Gate-tuning of graphene plasmons revealed by infrared nano-imaging. Nature 487, 82-85 (2012).

12. Sanyal, B., Eriksson, O., Jansson, U. & Grennberg, H. Molecular adsorption in graphene with divacancy defects. Phys. Rev. B 79, 113409 (2009).

13. An, J. et al. Domain (grain) boundaries and evidence of "twinlike" structures in chemically vapor deposited grown graphene. ACS Nano 5, 2433-2439 (2011).

14. Das, A., Chakraborty, B. & Sood, A. K. Raman spectroscopy of graphene on different substrates and influence of defects. Bull. Mater. Sci. 31, 579-584 (2008).

15. Kim, D. C. et al. The structural and electrical evolution of graphene by oxygen plasma-induced disorder. Nanotechnology 20, 375703 (2009).

16. Radchenko, T. M., Shylau, A. A. & Zozoulenko, I. V. Effect of charged line defects on conductivity in graphene: numerical Kubo and analytical Boltzmann approaches. Phys. Rev. B 87, 195448 (2013).

17. Fogler, M. M. In Preparation.

18. Yu, Q. et al. Control and characterization of individual grains and grain boundaries in graphene grown by chemical vapour deposition. Nature Mater. 10, 443-449 (2011).

19. Song, H. S. et al. Origin of the relatively low transport mobility of graphene grown through chemical vapor deposition. Sci. Rep. 2, 337 (2012).

20. Koepke, J. C. et al. Atomic-scale evidence for potential barriers and strong carrier scattering at graphene grain boundaries: a scanning tunneling microscopy study. ACS Nano. 7, 75-86 (2013).

21. Tapaszto, L. et al. Mapping the electronic properties of individual graphene grain boundaries. Appl. Phys. Lett. 100, 053114 (2012).